# Broaden the search for dark matter

**Bold strategies are needed to identify the elusive particles that should make up most of the universe's mass, say Mario Livio and Joe Silk.**

Dark matter is living up to its name. In spite of decades of compelling evidence from astronomical observations that matter that neither emits nor absorbs electromagnetic radiation exists, all attempts to detect dark matter's constituents so far have failed.

The presence of dark matter is inferred from its gravitational effects. Stars and gas clouds in galaxies and galaxies in clusters move faster than can be explained by the pull of visible matter alone. Light from distant objects may be distorted by the gravity of intervening dark material. The pattern of large-scale structures across the universe is largely dictated by dark matter. In fact about 85% of the universe's mass is dark, accounting for about a quarter of the total cosmic energy budget.

Despite its ubiquity, the nature of dark matter eludes us. Negative results have flowed from searches for candidate particles to explain it. In 2013, the Large Underground Xenon (LUX) experiment in South Dakota – the most sensitive detector to date – reported no signs of dark matter in its first 3 months of operation (1). The Large Hadron Collider (LHC) in Geneva, Switzerland, has found no evidence for the existence of supersymmetric particles – theoretically predicted partners to the known elementary particles – that some think are the most likely culprits.

Is there light at the end of this dark tunnel? Possibly – but only if searches become bolder and broader. More varied particle types should be sought. Definitive tests need to be devised to rule out some classes of dark matter and some theories. If dark matter remains undiscovered in the next decade, then physicists will have to seriously consider re-evaluating alternative theories of gravity.

*Exotic particles*

We know a little about dark matter (2). It cannot be made of baryons – particles of ordinary matter, including protons and neutrons, which are comprised of 3 quarks – because it does not absorb light or interact with electromagnetic waves. To avoid upsetting Big Bang nucleosynthesis, which successfully predicts the abundances of light elements such as deuterium, helium and lithium arising from interactions in the early universe, dark matter must lie beyond the 'standard model' of particle physics.

The main constituents are expected to be 'weakly interacting massive particles' (WIMPs)(3): with masses of a few tens to thousands of times that of the proton, that interact among themselves and with ordinary matter gravitationally and via the weak force, but not electromagnetically or via the strong nuclear force. To explain the way in which galaxies form and cluster, dark matter particles should be relatively slow moving, or 'cold'. If they were faster, and could have travelled beyond the typical scale of a protogalaxy, structures beyond the densest regions would have been washed out.

What kind of particles might WIMPs be? Many physicists guess the most likely candidates are the lightest of the supersymmetric (SUSY) particles. Theories of the early universe contain mathematical symmetries that allow for each particle we know, such as the electron, photon or quark, to have a (yet undiscovered) massive partner. Just after the Big Bang, almost all of the heavy SUSY particles decayed or annihilated with one another but the lightest, unable to decay further, were stable and survived.

The number of supersymmetric particles remaining depended on their masses and interaction strengths, which can be predicted from theory. If they interact too weakly, there would be too many; if too strongly, too few. The properties must be just right for the relic particle density to match that of the observed dark matter. Light SUSY particles fit that bill, motivating a plethora of experiments aiming to detect them. Yet no evidence for WIMPs has been seen.

Finding new particles is challenging. It took 4 decades of trials to find the Higgs boson. But through a series of current and planned experiments we should be able to rule out many candidates for dark matter in the next decade. The goal is to detect the particles that constitute the massive halo of dark matter that surrounds the Milky Way, as they pass through our detectors at rates of a few per second per square meter (4). But as they interact so weakly, a huge effort is needed to capture them. If we are lucky, a 300-day run of LUX, scheduled to begin later this year, could see WIMPs. But it may take bigger detectors – and more than one method.

Some estimates suggest that detectors a hundred times more massive than LUX will be needed to see WIMPs at the rates expected rates. LUX ZEPLIN, a planned upgrade that could begin operation around 2019, would use 7 tonnes of liquid xenon compared with LUX's 350 kilograms. We may yet require 100-tonne detectors, but that is the ultimate practical limit; further sensitivity is thwarted by an irreducible neutrino background, mainly from supernovae, the Sun and cosmic rays hitting the Earth's atmosphere.

Dark matter particles might be created in colliders. The LHC is expected to resume operation in 2015 at an energy of 14 TeV, twice that which led to the discovery of the Higgs boson. Supersymmetric particles and signatures of departure from the standard model might be glimpsed at these energies. But the lack of any SUSY signals to date suggests that we may need to push to much higher energies to see them. The 100-TeV collider, which many particle physicists support as the next step after LHC (construction could begin around 2020), would be exciting for dark matter searches (5).

Other dark matter experiments have made intriguing detections, whose interpretations are still being debated. DAMA/LIBRA at the Laboratori Nazionali in Gran Sasso, Italy, has been looking for changes in the flux of dark matter particles hitting the Earth for 14 years. As Earth orbits the Sun and as both travel through the Milky Way, the velocities of Earth and the Sun combine. For half the year, they are in the same direction, while in the other they are opposed. This produces an annual modulation in the rate at which the dark matter particles, travelling in random orbits, fall on the detector.

In 2013, DAMA/LIBRA reported such an annual variation with high statistical significance (more than 9$\sigma$) (6). Earlier this year, similar cycles, consistent in phase but with a larger amplitude than expected, were seen independently in 3 years of data from a US experiment, CoGeNT, but with relatively low statistical significance. Yet most physicists query whether the DAMA/LIBRA results really are due to WIMPs and not some other annual phenomenon, such as neutrons leached from the rocks surrounding the underground experiment in response to seasonal temperature variations.

Indirect attempts to detect dark matter have been equally inconclusive (8). The Alpha Magnetic Spectrometer (AMS-02) on board the International Space Station reported last year (9) an excess of positrons in the cosmic ray spectrum up to 350 GeV, consistent with being produced by dark matter particles colliding and annihilating. These results strengthened a similar 2008 claim from the space-based Payload for Antimatter/Matter Exploration and Light-nuclei Astrophysics (PAMELA) (10). But the positrons might have other origins, such as winds from pulsars (rapidly rotating neutron stars). Observations at higher energy by AMS-02 in the next 2 years might distinguish between these hypotheses.

Another source of excitement was the detection last year with the Fermi gamma-ray space telescope of an excess of gamma rays near the Galactic Center, where dark matter should concentrate. The narrow spectral line at 130 GeV apparently seen could indicate dark matter particle annihilations or decays. But a similar signal from the Earth's limb implies that at least part of the signal must be instrumental in origin. A conclusive test may come in the next couple of years from the HESS (High Energy Stereoscopic System) gamma-ray telescope in Namibia, which is observing the inner galaxy in the 100 GeV to 1 TeV energy range.

Perhaps more promising is the confirmation of an excess of gamma rays towards the Galactic Center by the Fermi satellite over the inner 10 degrees and in the 1-3 GeV energy range (11). After careful subtraction of various galactic foregrounds, the only rival explanation advocated other than dark matter annihilations, that of electron-positron pair production in outflows from millisecond pulsars, is argued to be incompatible with the observed gamma ray morphology, which matches the projected line-of-sight integral of the square of the dark matter density. The preferred particle mass is sensitive to modeling of electron propagation and lies in the range 10-40 GeV (12).

This is a mass regime that is also probed by the cosmic microwave background temperature fluctuations, which would be damped, perhaps excessively, by any delay in recombination, as expected for annihilations by dark matter particles in this mass range. The lower the particle mass, the more ionizing photons are produced for any cosmologically specified dark matter density. It is anticipated that the Planck satellite, with improved constraints on recombination via polarization measurements, will soon set a more definitive constraint on self-annihilating WIMPs below 30-40 GeV (13). At higher masses, direct detection becomes increasingly constraining.

The null results from LUX, the LHC and many other experiments are narrowing the range of possible particles that could explain dark matter. As searches become increasingly murky and claimed detections pop up only to disappear, physicists are becoming justifiably skeptical.

Some theorists have even started to wonder whether dark matter exists. Since the 1980s, a few have proposed modifying General Relativity to do away with the need for dark matter. Such radical ideas are being increasingly invoked to address another grave astrophysics problem, the origin of the 'dark energy' that accelerates the expansion of the universe. Most researchers think that we are far from needing new physical laws, especially since experimental avenues are still open. But unpleasant surprises are always possible.

There are two worst case scenarios. First, dark matter may not comprise one type of particle – as most current searches suppose – but many. Second, the particles may, like the gravitino (the SUSY counterpart of the graviton), interact only gravitationally, and be practically invisible to conventional detectors.

*New directions*

Existing experiments should be allowed to run their course. But new approaches will also be needed to close the window for dark matter particles in the next decade.

A dark matter modulation experiment like DAMA/LIBRA or CoGeNT in the southern hemisphere would gauge the extent of seasonal effects, which would be out of phase relative to the north.

Clumps or streams of dark matter moving through the Milky Way, which distort the rate at which particles impact the detectors, should be visible as disturbances in the motions of about a billion nearby stars tracked by the newly launched GAIA satellite, during its five-year mission.

At the LHC and next generation accelerators, particle collisions with missing energy – drawn by an unknown particle – or other unexpected signatures may illuminate the dark sector. But we must also broaden directed searches and exploit astrophysical methods.

First, we should look toward more massive particles, such as the SUSY particles. It will be difficult to detect heavy particles directly because there will be fewer of them. But gamma-ray astronomy may come to the rescue. The Cerenkov Telescope Array (CTA) – an international proposal to build over 100 telescopes to capture light flashes from gamma rays scattered by the atmosphere – would open the window after 2015 to 100-TeV energies. This energy coincides nicely with the highest limit on the WIMP mass expected from fundamental physics arguments. Such particles would generate TeV gamma rays when they annihilate or decay.

Second, broader categories of dark matter particles should be sought. Like ordinary matter, dark matter could be complex, perhaps carrying a tiny charge, or having internal states, like the electron levels of an atom. Changes in the Sun's internal oscillations as clouds of 'millicharged' particles scatter off electrons in the solar plasma might be detectable through helioseismology. The more spherical shapes of dark haloes of distant galaxies, expected if the particles interact electromagnetically, could be measured through gravitational lensing.

Third, there is the sterile neutrino. Released from thermal production constraints, which otherwise require the sum of neutrino masses to not exceed a few eV in order to avoid overly suppressing fluctuation power on cluster scales, where it would act as a hot dark matter component, the sterile neutrino mass can be targeted at the scale of dwarf galaxies. Here it resembles what has been dubbed warm dark matter and keV mass sterile neutrinos can help resolve three issues in the standard cold dark matter model. An excess of dark-matter dominated dwarfs is predicted, the center profiles are peaked and not cored as observed, and too many massive dwarfs are generated, too massive to be destroyed by the simplest feedback models. Until recently, the consensus on sterile neutrinos has been negative: for example, suppression of enough small-scale power to produce massive dwarf cores would require a 2 keV mass for the neutrino, but this would result in too late an epoch for galaxy formation in any credible hierarchical model (14).

This perspective has recently changed with the tentative discovery of a unidentified 3.5 keV x-ray spectral line from dark matter-dominated galaxy clusters. Could this line, admittedly not seen with high significance, be due to dark matter? The most attractive dark matter candidate to date is the two-photon decay of a sterile neutrino of mass 7 keV, and the decay probability can be tuned so that dark matter consists entirely of these neutrinos. Moreover there is an appealing coincidence. Ordinarily, 7 keV neutrinos are indistinguishable from CDM for structure formation. But for a particular production mechanism, resonant production of sterile neutrinos, one can have a nonthermal distribution of velocities whereby a 7 keV sterile neutrino affects matter as would a thermal 2 keV sterile neutrino. So one can one's cake and eat it too: galaxy formation occurs early enough, but the free-streaming effects remain important (15).

Fourth is the axion. Predicted to explain an anomaly in Quantum Chromo-Dynamics (QCD; the theory of the strong force), the electromagnetic signatures of axions have been long sought in the lab, without success. But string theory suggests that new types of ultralight axions may exist that would be slightly more 'warm' than cold dark matter. Mixes of cold and warm dark matter might explain why there are fewer dwarf galaxies than cold dark matter scenarios predict.

Astrophysicists might look for novel signals in old stars, such as neutron stars and white dwarfs. As stars orbit the galaxy, WIMPs accumulate. Collected in the core of a neutron star WIMPs might form a tiny black hole that could eventually devour the star, causing a violent explosion that has not yet been seen. The effect of WIMPs on the Sun's temperature profile could also be probed by helioseismology.

To refine theoretical and experimental strategies, particle physicists and astrophysicists need enhanced dialogues. The exciting prospect is that there is a relatively limited window to be explored, bounded at low masses by our failure to see anything, and at high masses by the limits of fundamental physics. A multidisciplinary approach to explore the 1 TeV to 100 TeV mass range should be the new frontier for the dark matter community.

Mario Livio is an astrophysicist at the Space Telescope Science Institute in Baltimore, Maryland. Joseph Silk is a professor at the Institut d'Astrophysique, Université Pierre et Marie Curie in Paris, and at the Department of Physics and Astronomy, the Johns Hopkins University, Baltimore, Maryland, and a Senior Fellow at the Beecroft Institute for Particle Astrophysics and Cosmology, University of Oxford.

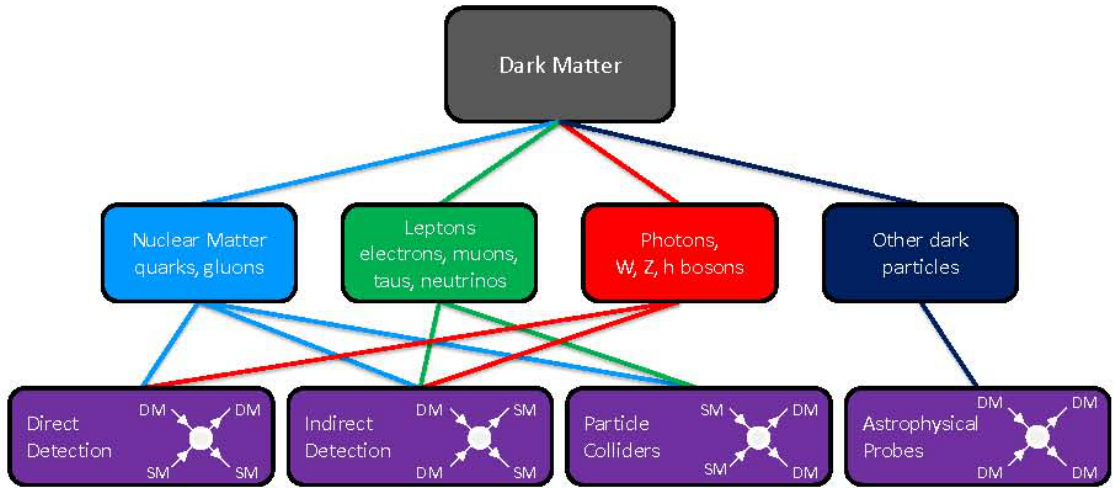

**Figure 1.** Dark matter may have non-gravitational interactions with one or more of four categories of particles: nuclear matter, leptons, photons and other bosons, and other dark particles. These interactions may then be probed by four complementary approaches: direct detection, indirect detection, particle colliders, and astrophysical probes. The lines connect the experimental approaches with the categories of particles that they most stringently probe. The diagrams give example reactions of dark matter (DM) with Standard Model particles (SM) for each experimental approach. From Bauer, D. et al. 2013, arXiv:1305.1605 [hep-ph].